\documentclass[colorlinks]{iopart}
\usepackage{bbm,iopams}

\usepackage{times,hyperref,tikz}

\usepackage{graphicx}
\usepackage{subfigure}

\usepackage{url}
\usepackage{cite}

\usepackage[normalem]{ulem}

\newcommand{\av}[1]{\overline{#1}} 
\newcommand{\xpct}[1]{\left\langle\smash{#1}\right\rangle}
\newcommand{\be}{\begin{equation}}
\newcommand{\ee}{\end{equation}}

\newcommand{\eqref}[1]{(\ref{#1})}

\newcommand{\vk}{\bi{k}}\newcommand{\vK}{\bi{K}}
\newcommand{\vp}{\bi{p}}
\newcommand{\vq}{\bi{q}}
\newcommand{\vr}{\bi{r}}
\newcommand{\vs}{\bi{s}}
\newcommand{\vu}{\bi{u}}

\newcommand{\gh}[1]{\hat\gamma_{#1}}		
\newcommand{\ghd}[1]{\hat\gamma^\dagger_{#1}}	%

\newcommand{\epn}[1]{\epsilon^0_{#1}}		
\newcommand{\ep}[1]{\epsilon_{#1}}

\newcommand{\matr}[4]{
\left(\begin{array}{cc}
#1 & #2 \\
#3 & #4 
\end{array}\right)
} 

\newcommand{\intdd}[1]{\int \frac{\rmd^d #1}{(2\pi)^d}}	

\newcommand{\calG}{\mathcal{G}}
\newcommand{\calV}{\mathcal{V}}

\newcommand{\ukp}{u_{\vk\vp}}
\newcommand{\vkp}{v_{\vk\vp}}
\newcommand{\ukpr}{u_{\vk\vp'}}
\newcommand{\vkpr}{v_{\vk\vp'}}

\newcommand{\nc}{n_{\rm c}}	
\newcommand{\Nc}{N_{\rm c}}	

\newcommand{\aIR}{\alpha} 

\newcommand{\avdeln}[2]{\av{\delta n}^{(#1)}_{#2}}
\newcommand{\avdn}[1]{\av{\delta n}{}^{(#1)}}

\begin{document}

\title[Bose-Einstein condensates in external potentials]
{Condensate deformation and quantum depletion of Bose-Einstein condensates in external potentials}

\author{C A M{\"u}ller$^1$ and C Gaul$^{2,3}$}

\address{$^1$ Centre for Quantum Technologies, National University of
  Singapore, Singapore 117543, Singapore}
\address{$^2$ GISC, Departamento de F\'isica de Materiales, Universidad
Complutense, E-28040 Madrid, Spain} 
\address{$^3$ CEI Campus Moncloa, UCM-UPM, Madrid, Spain} 

\pacs{03.75.Hh Static properties of condensates; thermodynamical,
  statistical, and structural properties.} 
%

\begin{abstract}
  The one-body density matrix of weakly interacting, condensed
  bosons in external potentials is calculated using inhomogeneous Bogoliubov theory. 
We determine the condensate deformation caused by weak external
potentials on the mean-field level. The momentum
distribution of quantum fluctuations around the deformed ground state 
is obtained analytically, and finally the resulting quantum
depletion is calculated. The depletion due to the external potential, 
or potential depletion for short, is a small correction to the
homogeneous depletion, validating our inhomogeneous Bogoliubov theory.
Analytical results are derived for weak lattices and spatially
correlated random potentials, with simple, universal results in the
Thomas-Fermi limit of very smooth potentials.

\end{abstract}


\section{Penrose-Onsager criterion and Bogoliubov theory for
  inhomogeneous condensates}
\label{intro} 

The phenomenon of Bose-Einstein condensation (BEC) 
plays a pivotal role in condensed-matter
physics. In recent years, the unique experimental possibilities
offered by dilute ultracold atomic gases have triggered a renewed
interest in Bose-Einstein
condensates.  
These are loaded into external potentials of a large variety, 
ranging from simple 
harmonic traps to increasingly complicated optical
lattices, all the way to random
potentials 
\cite{Lewenstein2007, 
Bloch2008, 
Sanchez-Palencia2010
}. 
The more complicated the confining potentials are, the greater
is the challenge to tell the condensate from the excitations, both
quantum and thermal. Yet, to be able to distinguish precisely between 
condensate and excitations under different circumstances 
is not only interesting from a conceptual
point of view. It is also important in order to understand the precise causal link between inhomogeneous BEC and certain physical properties, such as superfluidity. 

A criterion for BEC that applies
to interacting as well as inhomogeneous systems was proposed in 1956 by Penrose and Onsager \cite{Penrose1956}
and has remained in vigor until today: BEC occurs whenever the
one-body density matrix
(OBDM) 
\be\label{spdm}
\rho(\vr,\vr') = \langle\hat\Psi^\dagger(\vr)\hat\Psi(\vr')\rangle
\ee
has (at least) one macroscopically occupied eigenmode. 
As stated very clearly by Penrose and Onsager in their original paper, \emph{only if} the system
is completely homogeneous (i.e.\ translation invariant under periodic
boundary conditions), \emph{then} condensation occurs into a
single momentum component, namely the state with wave vector $\vk=0$ in the condensate
rest frame.  
Conversely, unless the additional assumption of spatial homogeneity is met, 
the zero-momentum occupation must not be used to determine the
condensate fraction. Recently, Astrakharchik and
Krutiksky \cite{Astrakharchik2011} have devised a quantum Monte Carlo
scheme of computing the OBDM and
condensate mode in external potentials, and Buchhold \textit{et al.}\ have 
studied collapse and revival of condensates under quenches of inhomogeneous lattices \cite{Buchhold2011}. 
In the present article, 
we employ analytical Bogoliubov theory, applicable to weakly interacting
condensates at low temperature, and calculate the condensate fraction and quantum
depletion of inhomogeneous Bose gases. 
  
Bogoliubov theory describes quantum fluctuations around a mean-field
condensate by splitting the quantum field 
\be\label{Bogoansatz}
\hat\Psi(\vr) = \Phi(\vr)
+\delta\hat\Psi(\vr)
\ee
 into a (large) mean-field condensate
$\Phi(\vr)$ and (small) quantum fluctuations $\delta\hat\Psi(\vr)$. In the symmetry-breaking picture,
the condensate $\Phi(\vr)=\xpct{\hat\Psi(\vr)}$ is the expectation
value of the quantum field. By consequence, $\xpct{\delta\hat\Psi(\vr)}=0$, and
the OBDM \eqref{spdm} splits into the sum of a condensed and non-condensed contribution, 
\be\label{OBDMBogo} 
\rho(\vr,\vr') = \Phi^*(\vr)\Phi(\vr') +
\xpct{\delta\hat\Psi^\dagger(\vr)\delta\hat\Psi(\vr')}.  
\ee
This form of the OBDM complies with the third
version of the Penrose-Onsager criterion \cite{Penrose1956}, often quoted in a
shortened manner. In full, this criterion reads as follows: 
If 
\be\label{PO3} 
|\rho(\vr,\vr')  - \Phi^*(\vr) \Phi(\vr') | \le n \gamma(|\vr-\vr'|)
\ee
with a function $\gamma(s)$ that is independent of the density $n=N/L^d$
and goes to zero at infinity, and if $\Phi(\vr)$ contains order
$N$ particles, then BEC occurs, and $\Phi(\vr)$ is a good
approximation to the condensate wave function. An OBDM with the
asymptotic form \eqref{PO3} is said to possess off-diagonal long-range
order. Bogoliubov's ansatz holds whenever the condensed
component is large, and then the
non-condensed component of the OBDM \eqref{OBDMBogo} can be bounded as
required by \eqref{PO3} \cite{Pitaevskii2003}. 

For further analysis, it is convenient to consider the bulk-averaged OBDM
\be\label{rhos}
\rho(\vs)= L^{-d}\int\rmd^d r
\xpct{\hat\Psi^\dagger(\vr)\hat\Psi(\vr+\vs)} = L^{-d} \sum_\vk
\rme^{\rmi\vk\cdot\vs} n_\vk
\ee
that depends only on a single vector $\vs$. It is the (inverse) Fourier
transform of the single-particle momentum distribution $n_\vk =
\xpct{\hat\Psi_\vk^\dagger\hat\Psi_\vk} $, where $\hat\Psi_\vk =
L^{-d/2} \int\rmd^d r \rme^{-\rmi\vk\cdot\vr} \hat\Psi(\vr)$ is the
particle annihilator in momentum representation. 
Under the Bogoliubov ansatz \eqref{Bogoansatz}, also the momentum
distribution splits into a condensed and non-condensed component: 
\be\label{nkdef}
n_\vk = |\Phi_\vk|^2 +
\xpct{\delta\hat\Psi^\dagger_\vk\delta\hat\Psi_\vk}\equiv n_{{\rm c}\vk} +
\delta n_\vk. 
\ee

Starting from these well-known concepts, we investigate in the
following the condensate deformation caused by weak external
potentials, calculate the corresponding fluctuation momentum
distribution, and finally determine  the resulting quantum
depletion. We apply our 
analytical theory to lattice potentials and spatially
correlated random potentials.  The paper is structured as follows. In Sec.~\ref{condef.sec}, we recall the momentum distribution and OBDM of the mean-field
condensate in presence of a weak external potential. 
In Sec.~\ref{momdis.sec}, we draw on the   
inhomogeneous Bogoliubov theory developed in
\cite{Gaul2011_bogoliubov_long} and derive the momentum
distribution of fluctuations in external potentials of arbitrary
form. 
Notably, we discover a universal
momentum distribution in the Thomas-Fermi regime of smoothly varying
potentials. In Sec.~\ref{condep.sec}, we compute the resulting quantum
depletion. The depletion caused by the
external potential is found to be a small correction proportional to the
depletion of the homogeneous Bose condensate, thus validating the
Bogoliubov theory of inhomogeneous condensates.  Section \ref{conclusion.sec}
concludes. 

\section{Condensate deformation}
\label{condef.sec}

\subsection{Gross-Pitaevskii theory} 

Within the Gross-Pitaevskii (GP) approach, the quantum fluctuations in
Eq.\ \eqref{Bogoansatz} are neglected, and it is relatively simple to
calculate the deformation of the condensate amplitude $\Phi(\vr)$ caused by an external potential $V(\vr)$.
One has to solve the stationary GP equation \cite{Pitaevskii2003} 
\be\label{GP} 
\left[(-\hbar^2\nabla^2/2m) +V(\vr) +g|\Phi(\vr)|^2 \right]\Phi(\vr) =
\mu\Phi(\vr)
\ee
at given chemical potential $\mu$. Numerically, the solution $\Phi(\vr)$ can be
computed rather efficiently by imaginary-time propagation of the
time-dependent GP equation \cite{Dalfovo1996}.

\subsection{Condensate deformation by a weak potential}

\begin{figure}[tbp]
 \subfigure[]{\includegraphics[angle=270,width=7.5cm]{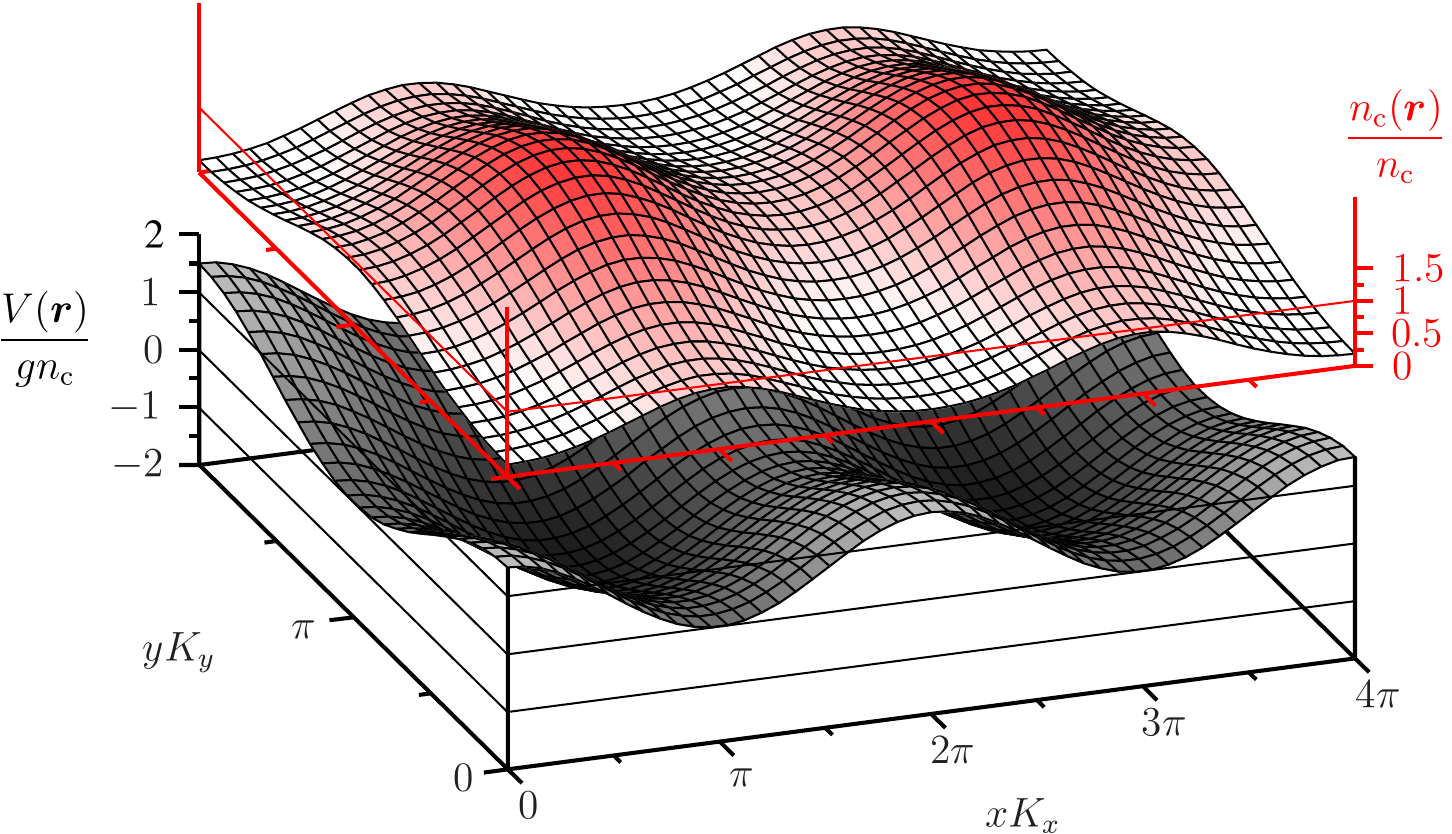}}\hfill
 \subfigure[]{\includegraphics[angle=270,width=6cm]{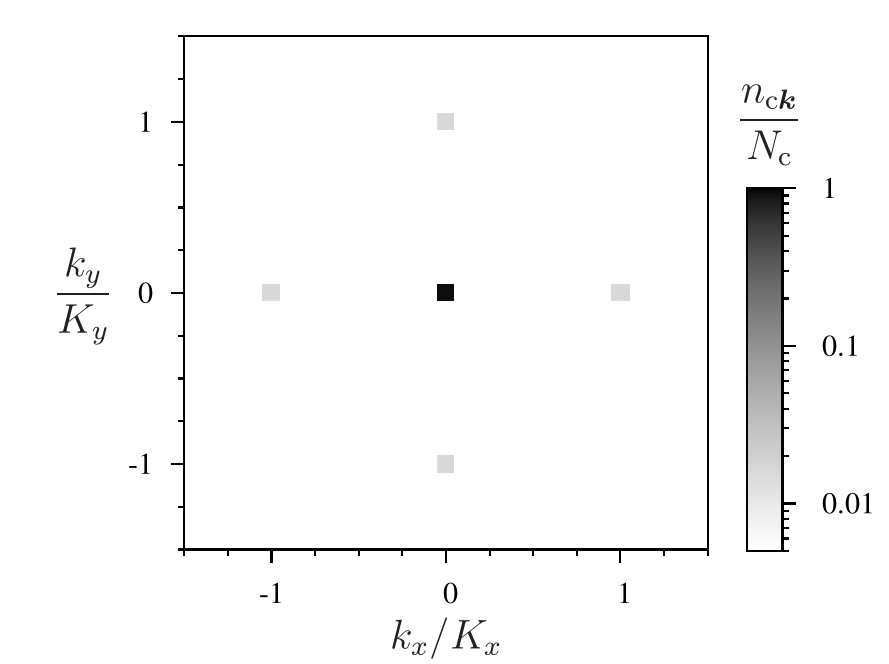}}
\caption{(a) Condensate density
  $\nc(\vr)/\nc$, perturbatively calculated via \eqref{Phik1} and \eqref{Phik2}, in a square lattice
  potential $V(\vr) =
V_x\cos (K_x x) +V_y \cos (K_y y)$ with $V_x=V_y=0.75gn_c$ and
$K_x\xi= K_y\xi=1$. 
The condensate adapts a deformed configuation in the external potential by avoiding peaks and
accumulating in wells. Due to the finite value of $K_j\xi$, the
  condensate profile is much smoother than the Thomas-Fermi profile 
$n_{\rm TF}(\vr)=\nc -V(\vr)/g$, which falls to zero around $\vr=0$
modulo the lattice period. 
(b) Corresponding condensate  momentum distribution $n_{{\rm c}
  \vk}/N_c$, Eq.~\eqref{nck}, showing $1/16\approx 6\%$ of
the total population in the 
$k$-components imprinted by the lattice. 
}\label{figLattice}
\end{figure}  

When the external potential is weak, it is straightforward to solve the GP equation perturbatively to the desired
order in powers of $V$ \cite{Sanchez-Palencia2006,Lugan2011}: $\Phi = \Phi^{(0)}
+ \Phi^{(1)} + \Phi^{(2)}+\dots$. In the following we work at fixed
average condensate density $\nc = L^{-d} \int\rmd^d r|\Phi(\vr)|^2$ and
adjust the chemical potential accordingly \cite{Gaul2011_bogoliubov_long}. 
In momentum representation, the
homogeneous condensate $\Phi^{(0)}_\vk = \Nc^{1/2}\delta_{\vk,0}$
receives the lowest-order deformations 
\begin{eqnarray}
\Phi_\vk^{(1)} & = -\Nc^{1/2} \tilde V_\vk , \label{Phik1}\\
 \Phi^{(2)}_\vk & = \frac{\Nc^{1/2}}{2 g \nc + \epn{k}} \sum_\vq \tilde V_{\vk-\vq}
   \tilde V_{\vq}\left[(1-\delta_{\vk0})\epn{q} -g \nc\right] \label{Phik2}.
\end{eqnarray} 
 The set of small parameters of this expansion are the reduced matrix elements  
\be\label{vtildek}
\tilde V_\vk = \frac{1-\delta_{\vk,0}}{\epn{k}+2 g \nc}  V_\vk,  
\ee
with $V_\vk = L^{-d}\int\rmd^d r \rme^{-\rmi\vk\cdot\vr} V(\vr)$ 
the bare potential matrix element, $\epn{k} = \hbar^2k^2/2m$ the bare kinetic energy, and 
$g \nc = \mu^{(0)}$ the chemical potential in absence of the external potential.
The condensate deformation  
follows the external potential only for $\epn{k}\ll g\nc$, or equivalently for wave vectors $k\xi \ll
1$ in terms of the healing length $\xi=\hbar/\sqrt{2m g\nc}$. 
Consequently, the condensate momentum distribution up to order $V^2$ becomes  
\be\label{nck}
n_{{\rm c}\vk} =  |\Phi_\vk|^2 = \Nc\left[(1-V_2) \delta_{\vk 0} + |\tilde
   V_\vk|^2\right] . 
\ee
Most particles in this distribution still have zero momentum, but
a small fraction 
\be\label{z2} 
V_2\equiv  \sum_\vk |\tilde V_\vk|^2 \ll1 ,
\ee 
of them have been promoted to finite 
momenta by the weak external potential. 

As an illustration, Fig.~\ref{figLattice} shows the condensate density $n_c(\vr)$
and its momentum distribution $n_{\mathrm{c}\vk}$, Eq.~\eqref{nck}, in
presence of a simple square lattice potential whose parameters are
such that $V_2 =1/16 = 6.25\%$. 
The external potential deforms the condensate, which tends to avoid 
potential peaks and accumulates in potential wells. The deformation is
periodic in real space, Fig.~\ref{figLattice}(a). In $k$-space, the
lattice shifts some population to the lattice momenta,
Fig.~\ref{figLattice}(b). To higher order in the 
external potential, also higher-order components would become visible
in the momentum distribution (cf.\ Fig.~2 in
\cite{Greiner2002}).

\subsection{Condensate deformation does not reduce mean-field condensate
  fraction} 

Condensate deformation is clearly a mean-field effect. 
The condensed part of the OBDM follows by inserting 
\eqref{nck}
into \eqref{rhos}: 
\be\label{rhoc}
\rho_{\rm c}(\vs) = \nc(1-V_2)+ \nc \sum_\vk \rme^{\rmi\vk\cdot\vs} |\tilde
V_\vk|^2. 
\ee 
By construction, $\rho_{\rm c}(0) = \nc$ is the total density of condensed
particles, which is kept fixed by adjusting the chemical
potential. Consequently, 
constant potential offsets have no effect, and thus $\tilde V_0=0$   
as implied by  \eqref{vtildek}.
Since $\tilde V_0=0$, one could be tempted to think that
the fluctuating part in \eqref{rhoc}  tends to zero in the
limit $s\to\infty$, which would imply that it does not contribute to
the off-diagonal long-range order. 
However, this reasoning is erroneous, as 
becomes quite evident in the case of a general lattice potential 
\be\label{latticer}
V(\vr) =
\sum_{j=1}^d V_j\cos(\vK_j\cdot\vr) . 
\ee
By consequence of the lattice periodicity, also the OBDM 
deformation is periodic in $\vs$, 
\be\label{fluct} 
\sum_\vk \rme^{\rmi\vk\cdot\vs} |\tilde
V_\vk|^2 = \frac{1}{2} \sum_{j=1}^d \tilde V_j^2 \cos(\vK_j\cdot\vs), 
\ee
with finite amplitudes $\tilde V_j =
V_j/(\epn{K_j}+2 g \nc)$. But a periodic deformation cannot
be bounded by a function $\gamma(s)$ that tends to
zero, and so it is the full momentum distribution \eqref{Phik2} that
describes the condensate,
in agreement with the Penrose-Onsager criterion.
Indeed, the (wrong) conclusion that only part of the field
$\Phi(\vr)$ constitutes the
condensate would contradict the mean-field ansatz used to
calculate it in the first place.  

If one performs additional configuration averages, then one accesses
no longer the full condensate fraction, but only its $\vk=0$ component. 
As an example, consider the 2D lattice potential \eqref{latticer} of
\fref{figLattice},  
and compute the angle-averaged OBDM 
$\av{\rho}(s) = (2\pi)^{-1} \int\rmd \theta \rho(\vs)$ \cite{Astrakharchik2011}. 
The inhomogeneous mean-field contribution \eqref{fluct} then averages to
$\frac{1}{2} \sum_j \tilde V_j^2 J_0(K_j s)$, with $J_0(.)$ the
Bessel function. This 
function indeed goes to zero as $s\to\infty$      
and thus yields, by construction, the 
translation-invariant component as a result. But working with an configuration-averaged OBDM does not 
prove that the zero-momentum component is a good indicator for
BEC, since the argument would be circular.  
Within mean-field theory, the condensate fraction is unity, regardless
of the specific form of the condensate mode $\Phi(\vr)$.

The same argument applies whenever an ensemble-average over a random
  potential distribution is performed. In fact, the  supposed disorder-depleted density
calculated in references  
\cite{Huang1992,Giorgini1994}
really is the averaged condensate deformation $n_c
\av{V_2}$ that follows from \eqref{z2} 
for a white-noise disorder potential.

Summarizing the mean-field discussion, we reiterate that
Bose-Einstein condensates form in many different spatial shapes and with 
different momentum distributions, determined by the competition of
kinetic, interaction, and potential energy. In a trap or any spatially
inhomogeneous potential,  the zero-momentum eigenstate relevant to free
space has no reason to determine the condensation into
a single-particle orbital. 
In short,  \emph{the population of momentum $\vk=0$ is not a good indicator for BEC in inhomogeneous
systems}. 

As a consequence, \emph{the population of momenta $\vk\neq 0$ does not
  measure the depletion of inhomogeneous condensates}, contrary to
what has been suggested, unfortunately,  in the
groundbreaking work of Huang and Meng \cite{Huang1992}, 
followed in this respect by numerous others
\cite{Astrakharchik2002,Giorgini1994,Kobayashi2002,Falco2007,Yukalov2007,Hu2009,Pilati2010,Krumnow2011}.
Within the Huang--Meng approximation
scheme, one cannot calculate the condensate
depletion induced by the external potential; this is achieved, for the
first time to our knowledge, in Sec.~\ref{condep.sec} below. To
  this end, we require a theory for
the fluctuations of inhomogeneous condensates. 

\section{Momentum distribution of quantum fluctuations}
\label{momdis.sec}

In this section, the momentum distribution $\delta n_\vk$ of
fluctuations, as defined by \eqref{nkdef}, is determined via
Bogoliubov theory as a function of the external
potential. For a weak potential, a perturbative, but fully analytical,
expression is obtained.  

\subsection{Bogoliubov excitations of an inhomogeneous condensate} 

Fluctuations around the inhomogeneous ground state $\Phi(\vr)$ are best
described using the density-phase representation: $\delta\hat\Psi(\vr) =  \exp\{\rmi \delta\hat\varphi(\vr)\}
[\Phi(\vr)^2+\delta\hat n(\vr)]^{1/2}-\Phi(\vr) $ develops as 
\be\label{Psir}
\delta \hat\Psi(\vr) = \frac{1}{2\nc}\check\Phi(\vr) \delta\hat
n(\vr) + \rmi \Phi(\vr) \delta\hat\varphi(\vr) +\dots 
\ee
Here, the inverse condensate amplitude $\check\Phi(\vr) = \nc/\Phi(\vr)$ is well
defined because weak external potentials do not fragment the condensate.
Likewise, highly excited states such as vortices are not considered,
and so $\Phi(\vr)>0$ holds everywhere. 

Fluctuations define Bogoliubov
quasi-particles, or ``bogolons''. Mathematically, these are  
obtained by the canonical transformation \cite{Giorgini1994}
\be\label{gammapdef}
\gh{\vp} =  
\delta \hat n_\vp /(2a_p\nc^{1/2}) + \rmi a_p \nc^{1/2}
\delta\hat\varphi_\vp. 
\ee
Here,  $a_p = (\epn{p}/\ep{p})^{1/2}$ for all $\vp\neq 0$ is the traditional Bogoliubov
transformation parameter, given by the ratio of free-particle
dispersion $\epn{p} = \hbar^2p^2/2m$ to the Bogoliubov dispersion
$\ep{p} = [\epn{p}(\epn{p} + 2 g \nc)]^{1/2}$. For the zero mode, it
is appropriate to define $a_0=1$, as discussed in \ref{Bogotrafo.apx}. 
Importantly, the fluctuations 
$\delta \hat n_\vp$ and $\delta\hat\varphi_\vp$ are the Fourier components of density and phase
deviations away from the deformed mean-field ground state $\Phi(\vr)$---and not
from the homogeneous background $\nc^{1/2}$ since then the deformation
effect of the potential would be missed.   By consequence of
\eqref{gammapdef}, 
the fluctuation \eqref{Psir} is expressed via bogolons as
\be\label{delPsikgamma}
\delta\hat\Psi_\vk = \sum_\vp \left(u_{\vk\vp}\gh{\vp} -
  v_{\vk\vp}\ghd{-\vp}\right),   
\ee 
where the inhomogeneous Bogoliubov transformation matrices
\begin{eqnarray}
u_{\vk\vp} & = \frac{1}{2\sqrt{\Nc}}\left[ a_p^{-1}\Phi_{\vk-\vp} +
    a_p\check\Phi_{\vk-\vp} \right], \label{ukp}\\
v_{\vk\vp} & =  \frac{1}{2\sqrt{\Nc}}\left[ a_p^{-1}\Phi_{\vk-\vp} -
    a_p\check\Phi_{\vk-\vp} \right],   \label{vkp}
\end{eqnarray}
contain the Fourier coefficients of the condensate
amplitude, $\Phi_\vk = L^{-d/2} \int\rmd^d r \rme^{-\rmi\vk\cdot\vr}
\Phi(\vr) $,  and its inverse, 
$\check\Phi_\vk = [\nc/\Phi]_\vk$. 
Some useful properties of this transformation to Bogoliubov
quasiparticles are discussed in \ref{Bogotrafo.apx}.

Inserting \eqref{delPsikgamma} and its Hermitian conjugate into
$\delta n_\vk = \xpct{\delta\hat\Psi^\dagger_\vk\delta\hat\Psi_\vk}$ 
brings the fluctuation momentum distribution in the form 
\be\label{deltank}
\fl \delta n_\vk = \sum_{\vp,\vp'}\left\{
\delta_{\vp\vp'} |\vkp|^2 
+\left(\ukp^*\ukpr + \vkp^*\vkpr \right) \xpct{\ghd{\vp}\gh{\vp'}} 
- (\ukp^*\vkpr\xpct{\ghd{\vp}\ghd{-\vp'}} +c.c.)
\right\} . 
\ee
This equation holds to arbitrary order in
potential strength $V$, as long as expansion
  \eqref{Psir} is valid, namely for a non-vanishing condensate
  amplitude $\Phi(\vr)$ and negligible higher-order fluctuations.  
In order to compute the expectation values $
\xpct{\ghd{\vp}\gh{\vp'}} $ 
and $\xpct{\ghd{\vp}\ghd{-\vp'}}$, we need
to specify the Hamiltonian of inhomogeneous Bogoliubov fluctuations.

\subsection{Inhomogeneous Bogoliubov Hamiltonian} 

The quadratic Hamiltonian for the Bogoliubov excitations of an
inhomogeneous Bose gas was derived in \cite{Gaul2011_bogoliubov_long,Gaul2008} by a
saddle-point expansion of the many-body Hamiltonian around the deformed ground-state solution
$\Phi(\vr)$: 
\be\label{eqInhomBgHamiltonian_Gamma}
\hat H = \sum_{\vk} \ep{k} \hat\Gamma_{\vk}^\dagger\hat\Gamma_{\vk} + \sum_{\vk,\vk'} 
\hat\Gamma^\dagger_{\vk}
\calV_{\vk \vk'}
\hat\Gamma_{\vk'}.  
\ee
The Bogoliubov-Nambu (BN) 
pseudo spinor $\hat\Gamma_{\vk}^\dagger=
(\ghd{\vk},\gh{-\vk})/\sqrt{2}$ allows for a rather compact
notation. The only approximation in the derivation of the Hamiltonian 
 is the neglect of third and fourth
order terms in the fluctuations. 
In contrast, \eqref{eqInhomBgHamiltonian_Gamma} is still exact in the external potential 
strength and has the structure $\hat H = \hat H^{(0)} + \hat
H^{(V)}$. The price to be paid for spatial inhomogeneity is the 
appearance of the 
effective scattering vertex  
\be\label{calV}
\calV_{\vk\vk'} =
\matr{W_{\vk\vk'}}{Y_{\vk\vk'}}{Y_{\vk\vk'}}{W_{\vk\vk'}}. 
\ee
Its matrix elements, 
\begin{eqnarray}
W_{\vk\vk'} &= \frac{1}{4} \left[a_k a_{k'}R_{\vk\vk'} +
  a_k^{-1}a_{k'}^{-1}S_{\vk\vk'}  \right], \label{Wkkpr}\\
Y_{\vk\vk'} &= \frac{1}{4} \left[a_k a_{k'}R_{\vk\vk'} -
  a_k^{-1}a_{k'}^{-1}S_{\vk\vk'}  \right], \label{Ykkpr}
\end{eqnarray} 
are entirely determined by  mean-field amplitudes $\Phi_\vk$ and $\check\Phi_\vk$ via 
\begin{eqnarray}
S_{\vk\vk'} &= \frac{2g}{L^{d}}\xi^2\vk\cdot\vk' (1-\delta_{\vk\vk'}) \sum_\vp \Phi_{\vk-\vp}\Phi_{\vp-\vk'}, \\
R_{\vk\vk'} &=  \frac{2g}{L^{d}}\xi^2 \sum_\vp\left[\vk\cdot\vk' +
  (\vk+\vk'-2\vp)^2 \right]\check\Phi_{\vk-\vp}\check\Phi_{\vp-\vk'} - 2\epn{k}\delta_{\vk\vk'} .
\end{eqnarray}
Here, we have dropped the superscripts $S^{(V)}$ and $R^{(V)}$ used in 
\cite{Gaul2011_bogoliubov_long}.

\subsection{Bogolon populations} 
\label{bogopop.sec} 

It is in principle possible to diagonalize the Hamiltonian \eqref{eqInhomBgHamiltonian_Gamma} numerically, for each
realization of the external potential, after having solved the nonlinear GP equation
\eqref{GP}.  However, for the purpose of analytical calculations
in weak external potentials, a more economic strategy is to 
calculate the bogolon populations required in \eqref{deltank}
perturbatively. We assume an 
equilibrium state at finite temperature $T$. The normal
expectation value can be expressed via the
single-quasiparticle Matsubara-Green (MG) function: 
\be\label{expectgdgMG.eq}
\xpct{\ghd{\vp'}\gh{\vp}} = -\lim_{\tau\to0^-} G_{\vp\vp'}(\tau) =
-\frac{1}{\beta}\sum_{n\in\mathbbm{Z}} G_{\vp\vp'}(i\omega_n)
\ee
where $\omega_n= 2\pi n/\beta$ are the bosonic Matsubara
frequencies related to the inverse temperature
$\beta=1/k_\mathrm{B}T$. 
Similarly, the anomalous expectation value
\be\label{expectgdgdMG.eq}
\xpct{\ghd{\vp'}\ghd{-\vp}}  =
-\frac{1}{\beta}\sum_{n\in\mathbbm{Z}} F_{\vp\vp'}(i\omega_n) 
\ee
is expressed in terms of the anomalous MG function $F(z)$. Together,
the normal and anomalous MG function enter the Nambu-MG matrix
$\calG = \footnotesize \matr{G}{F^\dagger}{F}{G^\dagger}$ which expands
as $\calG = \calG_0 + \calG_0 \calV \calG_0 + \calG_0
\calV \calG_0 \calV \calG_0 + \dots $ 
The free
propagator is diagonal in Nambu space and momentum representation, $\calG_{0\vp}(i\omega_n) = \mathrm{diag}(G_{0\vp}(i\omega_n),  
G_{0\vp}(-i\omega_n))$ with $G_{0\vp}(z)= [z-\epsilon_p]^{-1}$. 
Matsubara sums like \eqref{expectgdgMG.eq} and \eqref{expectgdgdMG.eq}
are carried out using textbook recipes such as (11.58) in \cite{Bruus2004};
each simple pole with energy $\epsilon_p$ contributes one Bose-Einstein occupation number
\begin{equation} 
\nu:=\nu(\beta\epsilon_p) = [\exp(\beta\epsilon_p)-1]^{-1}.
\end{equation}  

The expectation values \eqref{expectgdgMG.eq}, \eqref{expectgdgdMG.eq}
are then straightforward to calculate. For
brevity, we 
present here only the diagonal result for $\vp'=\vp$, i.e.\ the bogolon population  $\nu_\vp\equiv
\xpct{\ghd{\vp}\gh{\vp}}$, up to order $\calV^2$:   
\begin{eqnarray} 
\nu_\vp = \nu + \frac{\partial \nu}{\partial \epsilon}
W_{\vp\vp} +  \sum _{\vp'} 
& \left\{ \frac{1}{\epsilon-\epsilon'}\left(\frac{\partial \nu}{\partial
      \epsilon} - \frac{\nu'-\nu}{\epsilon'-\epsilon}\right)
W_{\vp\vp'} W_{\vp'\vp}    \right. \nonumber  \\ 
& - \left. \frac{1}{\epsilon+\epsilon'}\left(\frac{\partial \nu}{\partial
      \epsilon} - \frac{1+\nu'+\nu}{\epsilon'+\epsilon}\right)
Y_{\vp\vp'} Y_{\vp'\vp} 
     \right\}  .
 \end{eqnarray}
Here the short-hand notations 
$\epsilon=\epsilon_p$, $\epsilon'= \epsilon_{p'}$, and $\nu'=\nu(\beta\epsilon')$ are used.  
At temperature $T=0$ when all occupation numbers and their derivatives
vanish, there is only a single
finite contribution to the normal bogolon population due to the
external potential, to order $\calV^2$: 
 \be\label{nu2k0.eq}
\nu_{\vp}  =  \sum _{\vp'}  \frac{1}{(\epsilon+\epsilon')^2} Y_{\vp\vp'} Y_{\vp'\vp} .
 \ee
The bogolon quasi-particles are populated by the random potential even at
zero temperature because the
full Hamiltonian \eqref{eqInhomBgHamiltonian_Gamma} is not diagonal in the 
basis that diagonalizes $H^{(0)}$. 

Similarly, the anomalous population
$\pi_\vp= - \xpct{\ghd{\vp} \ghd{-\vp}}$ reads, to order $\calV^2$,  
\begin{eqnarray}  
\pi_\vp =  \frac{1+2\nu}{2\epsilon} Y_{\vp\vp} -
  \frac{1}{\epsilon}  \sum _{\vp'} 
 \frac{\epsilon- \epsilon'
   +2(\epsilon\nu'-\epsilon'\nu)}{(\epsilon-\epsilon')(\epsilon+\epsilon')} Y_{\vp\vp'} W_{\vp'\vp}. 
\end{eqnarray} 
At $T=0$, the result simplifies slightly: 
\be \label{pi2k0.eq}
\pi_{\vp}  = \frac{1}{2\epsilon} Y_{\vp\vp}
- \frac{1}{\epsilon}  \sum _{\vp'} 
 \frac{1}{\epsilon+\epsilon'} Y_{\vp\vp'} W_{\vp'\vp} .
\ee  
With this, everything is in place to calculate the full momentum
distribution \eqref{deltank}, or equivalently the OBDM \eqref{rhos}. 
In the following, we pursue a fully analytical
calculation by a perturbative expansion up to order $V^2$ in the bare
external potential. In the upcoming section \ref{weakpot.sec} 
the Bogoliubov transformation matrices  \eqref{ukp} and
\eqref{vkp} are determined together with the scattering
matrix elements \eqref{Wkkpr} and \eqref{Ykkpr}. In Sec.~\ref{momdisv2.sec} the  results are 
collected into a compact expression for the momentum distribution. 

\subsection{Weak-potential expansion} 
\label{weakpot.sec}

The perturbation expansion \eqref{Phik1}--\eqref{Phik2}  
of the
condensate amplitude $\Phi_\vk$
in powers of $V$ 
implies a similar expansion 
for the Bogoliubov transformation matrices  \eqref{ukp} and \eqref{vkp}:
\begin{eqnarray}
\ukp = \ukp^{(0)} +  \ukp^{(1)} +    \ukp^{(2)} +\dots, \\ 
\vkp = \vkp^{(0)} +  \vkp^{(1)} +    \vkp^{(2)} +\dots
\end{eqnarray} 
To zeroth order in the external potential, the transformation matrices
are diagonal in momentum, as required by translation invariance, and
the traditional Bogoliubov amplitudes are recovered: 
\begin{eqnarray}
\ukp^{(0)} = \frac{1}{2}(a_p^{-1}+a_p)\delta_{\vk\vp} \equiv u_p\delta_{\vk\vp},  \\ 
\vkp^{(0)} =   \frac{1}{2}(a_p^{-1}-a_p)\delta_{\vk\vp} \equiv v_p\delta_{\vk\vp}. 
\end{eqnarray} 
To first order, the matrix elements are
proportional to the potential matrix element \eqref{vtildek}: 
\begin{eqnarray}
\ukp^{(1)} = -v_p\tilde V_{\vk-\vp},  \qquad
\vkp^{(1)} = -u_p\tilde V_{\vk-\vp}.
\end{eqnarray} 
For the second-order matrices, only  the diagonal matrix
elements will be required: 
 \begin{eqnarray}
u_{\vk\vk}^{(2)} = \frac{u_k-2v_k}{2} V_2,  \qquad 
v_{\vk\vk}^{(2)} = \frac{v_k-2u_k}{2} V_2,
\end{eqnarray} 
where $V_2$ of Eq.~\eqref{z2} is second order in $V$. 

 As spelled out in
\cite{Gaul2011_bogoliubov_long}, also the BN scattering potential
\eqref{calV} admits the expansion  
$ \calV  = \calV^{(1)} +  \calV^{(2)} + \dots $. 
Also here, the first-order scattering amplitudes 
\begin{eqnarray} 
W^{(1)}_{\vk \vp} = g \nc \tilde w^{(1)}_{\vk \vp} \tilde V_{\vk-\vp},
\qquad 
Y^{(1)}_{\vk \vp} =g \nc \tilde y^{(1)}_{\vk \vp}  \tilde V_{\vk-\vp}
\end{eqnarray}  
are proportional to \eqref{vtildek}, with 
\begin{eqnarray}
\tilde w^{(1)}_{\vk \vp} &= 
\xi^2 \left[a_k a_p( k^2 + p^2 -\vk\cdot\vp) 
- {a_k^{-1} a_p^{-1}} \vk\cdot\vp  \right]  \nonumber\\
& = \xi^2 \left[(u_k-v_k) (u_p-v_p)(k^2+p^2) - 2
  (u_k u_p+v_k v_p)\vk\cdot\vp \right], \label{tildew1}  \\ 
\tilde y^{(1)}_{\vk \vp} &=  \xi^2 \left[ a_k a_p ( k^2 + p^2 -\vk\cdot\vp
) + a_k^{-1} a_p^{-1} \vk\cdot\vp  \right]  \nonumber\\
& = \xi^2 \left[(u_k-v_k) (u_p-v_p)(k^2+p^2) + 2
  (u_k v_p+v_k u_p)\vk\cdot\vp \right]. \label{tildey1} 
\end{eqnarray}
Second-order scattering amplitudes are later only needed  for 
$\vk=\vp$: 
\be
W^{(2)}_{\vk \vk} = Y^{(2)}_{\vk \vk} 
 = g \nc \sum_{\vp}   \tilde y^{(2)}_{\vk \vp}  |\tilde V_{\vk-\vp}|^2
\ee 
with   
\begin{equation} 
\tilde y^{(2)}_{\vk\vp}  = 2 a_k^2\xi^2 [k^2+(\vk-\vp)^2].
\label{tildey2}
\end{equation}

\subsection{Momentum distribution} 
\label{momdisv2.sec} 

Collecting all results up to order $V^2$, the 
single-particle fluctuation momentum distribution \eqref{deltank} reads 
\be\label{deltank02}
\delta n_\vk  = \xpct{\delta\hat\Psi^\dagger_\vk\delta\hat\Psi_\vk} =
\delta n_\vk^{(0)} + \delta n_\vk^{(2)},    
\ee 
where the superscript indicates the order in the external potential
strength $V$. To zeroth order, i.e.\ for the homogeneous
system, we recover the well known  zero-temperature 
momentum distribution \cite{Pitaevskii2003} 
\be\label{delnk0}
\delta n_\vk^{(0)} = v_k^2 = \frac{(a_k-a_k^{-1})^2}{4} = \frac{(\ep{k}-
  \epn{k})^2}{4\ep{k}\epn{k}} =
\frac{1+(k\xi)^2}{2k\xi\sqrt{2+(k\xi)^2}} - \frac{1}{2}
\ee 
as a consequence of the two-body contact interaction. 
The first-order term vanishes  by momentum
conservation,  $\delta n_\vk^{(1)}=0$.  The external potential induces
the second-order shift  
\be\label{deltan2MV}
\delta n_\vk^{(2)} =
\sum_\vp\tilde  M^{(2)}_{\vk\vp} |\tilde V_{\vk-\vp}|^2, 
\ee
whose kernel is defined in terms of \eqref{tildew1}, \eqref{tildey1},
and \eqref{tildey2}:
\begin{eqnarray}\label{kernelM} 
 \tilde M^{(2)}_{\vk\vp} & =  (v_k^2 - 2u_k v_k + u_p^2) - 2(u_k u_p+ v_k v_p)
\frac{g \nc\tilde y^{(1)}_{\vk\vp}}{\ep{k}+\ep{p}}  \nonumber\\ 
&+ (u_k^2+v_k^2) \frac{(g \nc\tilde y^{(1)}_{\vk\vp})^2}{(\ep{k}+\ep{p})^2}
+ \frac{ g \nc u_k v_k}{\ep{k}} \left\{ \tilde y^{(2)}_{\vk\vp} -
  \frac{2 g \nc}{\ep{k}+\ep{p}}\tilde
    y^{(1)}_{\vk\vp} \tilde w^{(1)}_{\vk\vp}\right\}. 
\end{eqnarray} 
This expression, together with the preceding general form \eqref{deltank},
constitutes the main result of this calculation.  
From here on, we explore its consequences 
by studying two generic 
examples of external potentials: a weak lattice potential on the one
hand, and a random potential on the other.

\subsubsection{Lattice potential --} A pure lattice potential like
\eqref{latticer} has only the Fourier components $V_\vk=
\frac{1}{2}\sum_{j=1}^d V_j\left(\delta_{\vk,\vK_j}
  +\delta_{\vk,-\vK_j} \right)$, such that  
\be\label{latticek}
|V_\vk|^2 = 
\frac{1}{4}\sum_{j=1}^d V_j^2\left(\delta_{\vk,\vK_j}
  +\delta_{\vk,-\vK_j} \right).
\ee
Thus, the momentum distribution shift \eqref{deltan2MV} is given by 
\be\label{latticedelnk}
\delta n_\vk^{(2)} = \frac{1}{4}\sum_{j=1}^d\frac{(V_j/g \nc)^2}{[2+(K_j\xi)^2]^2}
\left(  \tilde M^{(2)}_{\vk\,\vk+\vK_j} + \tilde
  M^{(2)}_{\vk\,\vk-\vK_j}\right). 
\ee

\begin{figure} 
\centerline{\includegraphics[angle=270,width=8cm]
{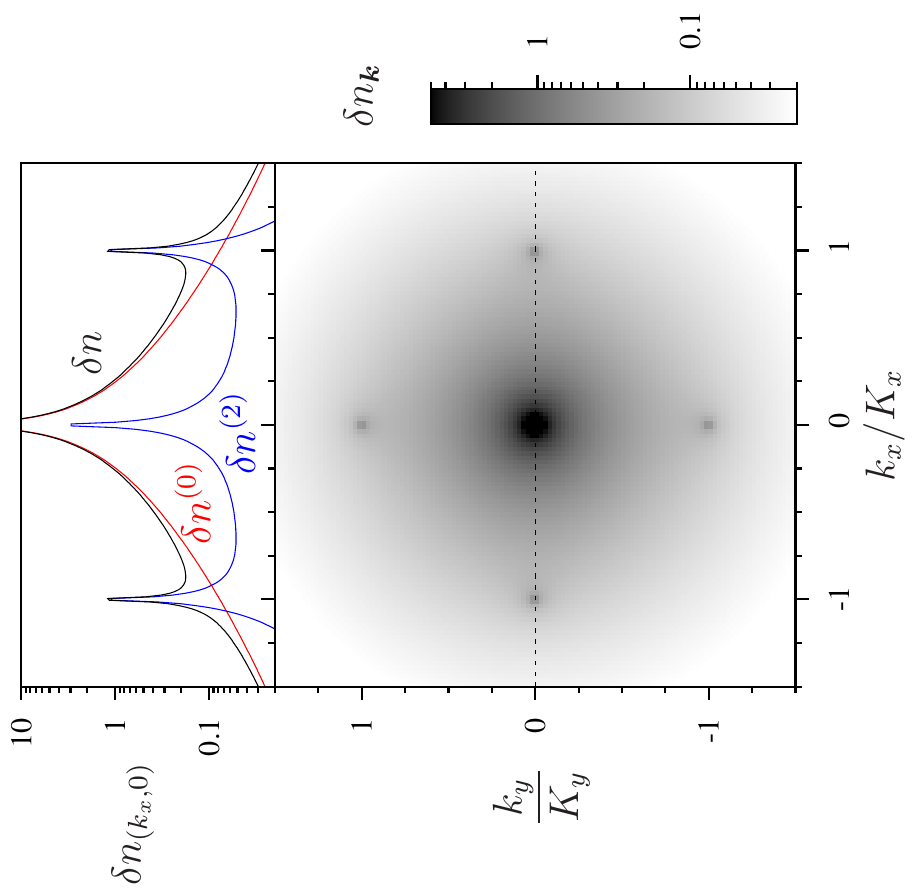}} 
\caption{Zero-temperature momentum distribution $\delta n_\vk$ of
  quantum fluctuations, Eq.~\eqref{deltank02}, in
  the square lattice potential of Fig.~\ref{figLattice}. 
The upper
  panel shows a cut along $k_y=0$.  The main contribution stems from the
isotropic background $\delta n^{(0)}_\vk$, Eq.~\eqref{delnk0}. The
potential-induced distribution $\delta n^{(2)}_\vk$,
Eq.~\eqref{latticedelnk}, reflects the lattice structure.  
}\label{latticefluctplot}
\end{figure}  

Figure \ref{latticefluctplot} shows the momentum distribution \eqref{deltank02} in the
square lattice potential of Fig.~\ref{figLattice}. The top panel
shows a cut through the distribution at $k_y=0$, together with the
separate contributions of the homogeneous
and potential-induced distribution.  
Compared to the mean-field  \fref{figLattice}, the quantum fluctuations
broaden the momentum distribution substantially. 

In formula  
\eqref{latticedelnk},
the product $K_j\xi$ compares the characteristic length scale of the
potential, $K_j^{-1}$,  with  the
condensate healing length $\xi$. For not-too-low densities and typical interaction strengths achievable with
ultracold atoms, one easily reaches $K_j\xi\ll1$, known as the Thomas-Fermi (TF) regime. The healing length is the
characteristic scale also for the entire kernel \eqref{kernelM}. In
the deep TF regime $K_j\xi\to 0$,
and for finite momenta $k>K_j$, this complicated kernel can be
approximated by the diagonal term $\tilde M_{\vk\vk}^{(2)}
=[(k\xi)^2-1]/\{k\xi[2+(k\xi)^2]^{5/2}\}$. The potential-induced change of momentum
distribution then takes the simple isotropic form
\be\label{delnkTF}
\delta n_{\vk\mathrm{TF}}^{(2)} = \frac{v^2}{4} \frac{(k\xi)^2
  -1}{k\xi [2+(k\xi)^2]^{5/2}}.
\ee
Here, $v^2=\frac{1}{2}\sum_j
(V_j/g \nc)^2$ measures the 
potential variance in units of the
mean-field interaction energy.  
In the TF regime \eqref{delnkTF}, the external potential is found to shift population from low momenta $k\xi<1$ to high
momenta $k\xi>1$, and this independently of the detailed form of the
potential.  

\subsubsection{Random potential --}

A random potential can be seen as 
a superposition of many lattices with a random distribution of Fourier
components $V_\vk$, specified by the ensemble averages  $\av{V_\vk}$, $\av{V_\vk V_\vp}$, etc.  Here, we assume without loss of generality that the
potential is centred, $\av{V(\vr)}=0$ or $V_0=0$.  All we need at
order $V^2$ then is the pair correlator  
\begin{equation}\label{eqDisorderCorrelator}
\av{V_\vq V_{-\vq'}} = L^{-d}\delta_{\vq \vq'} V^2 \sigma^d  C_d(\vq\sigma).
\end{equation}
The dimensionless function $C_d(\vq\sigma)$
characterizes the potential correlation on the length
scale $\sigma$; the normalization is chosen such that
$(\sigma/L)^d\sum_\vq C_d(\vq\sigma) = 1$ in the thermodynamic
limit. Using
\eqref{eqDisorderCorrelator}, the ensemble-averaged change of the
single-particle momentum distribution \eqref{deltan2MV} takes the form 
\be\label{avdelnk}
\avdeln{2}{\vk} = v^2 \frac{\sigma^d}{L^d} \sum_\vq \tilde
M^{(2)}_{\vk\,\vk-\vq}  \frac{ C_d(\vq\sigma)}{[2+(q\xi)^2]^2}
\ee
where $v^2 = (V/g \nc)^2$ is the potential variance in units of
mean-field interaction energy. 

\begin{figure}
\hfill
\includegraphics{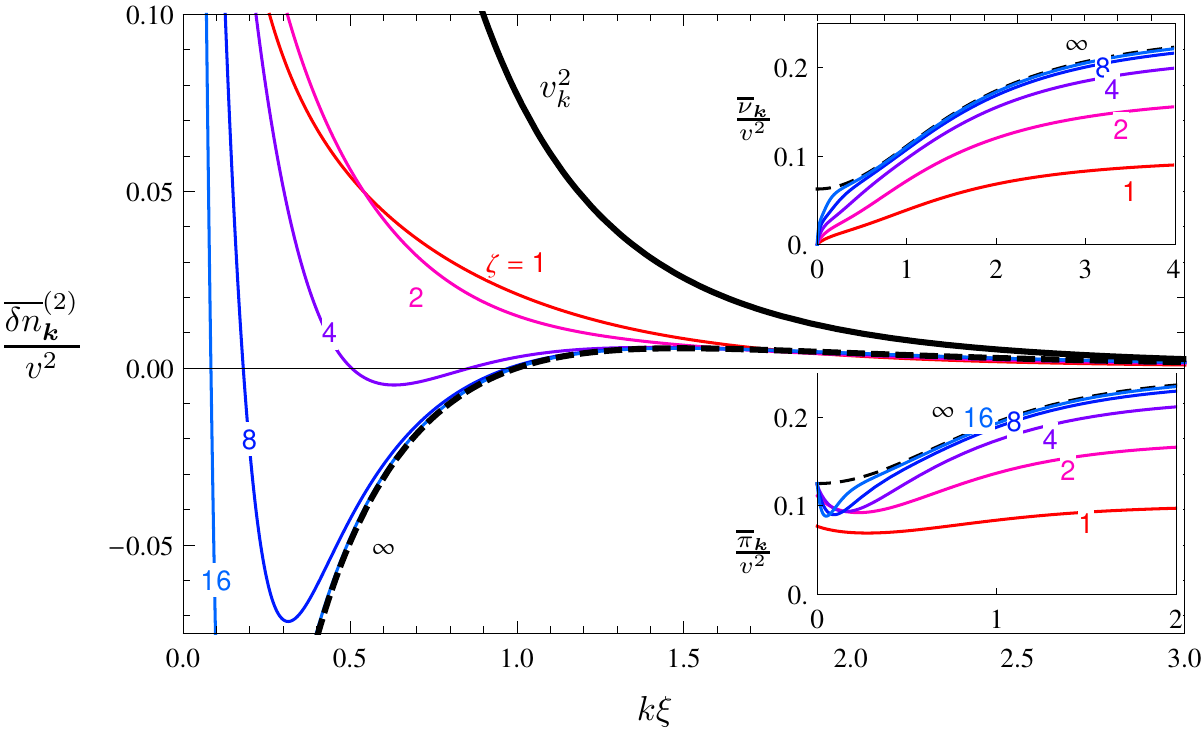}
\caption{Quantum-fluctuation momentum distribution shift \eqref{avdelnk} induced by a
  2D random potential of reduced variance $v^2 = V^2/(g \nc)^2$ and
  Gaussian correlation, \eqref{gausscorr.eq}, for different values of the
  correlation length relative to the condensate healing length, $\zeta
  = \sigma/\xi$. For comparison, also the momentum distribution
  $\delta n^{(0)}_k= v_k^2$, Eq.~\eqref{delnk0}, of the 
  homogeneous fluctuations is
  shown. The insets show the normal and anomalous
  populations of Bogoliubov excitations, \eqref{avnuk} and \eqref{avpik},
  respectively, also as function of $k\xi$.  In the TF limit $\zeta\to\infty$, all 
  distributions converge towards the universal expressions 
  \eqref{delnkTF}, \eqref{avnu2k0TF.eq}, and \eqref{avpi2k0TF.eq},
  respectively, shown in dashed black.}
\label{nk2.fig} 
\end{figure} 

Figure \ref{nk2.fig} shows the ensemble-averaged, isotropic, momentum
distribution \eqref{avdelnk} that is induced
by a random
potential with Gaussian correlation 
\begin{equation} \label{gausscorr.eq}
C_d(q\sigma) = (2\pi)^{d/2} \exp\{-q^2\sigma^2/2\}
\end{equation} 
in $d=2$ dimensions  as function of $|\vk|\xi$. Different curves
correspond to different values of the
  correlation parameter  $\zeta
  = \sigma/\xi$, namely the correlation length relative to the
  condensate healing length. 
Just as for the lattice, also here things simplify considerably in the
TF regime $\sigma\gg\xi$ where the disorder correlation length is much longer than
the condensate healing length. Then, the potential correlator tends to
a $\delta$-distribution, and \eqref{avdelnk} reduces to
\eqref{delnkTF}, plotted in dashed black. 
Clearly, the momentum distribution of quantum
fluctuations is given by the universal form
\eqref{delnkTF}  in any external potential that is sufficiently smooth to 
yield a Thomas-Fermi condensate profile. 

The insets of Fig.~\ref{nk2.fig} show the ensemble-averaged, normal
and anomalous bogolon 
populations 
$\nu_\vk = \xpct{\ghd{\vk} \gh{\vk}}$ and 
$\pi_\vk= - \xpct{\ghd{\vk} \ghd{-\vk}}$  at zero temperature.  
These populations, 
\eqref{nu2k0.eq} and \eqref{pi2k0.eq}, 
are given by 
\begin{eqnarray}
\av{\nu}_{\vk}  & = V^2 \frac{\sigma^d}{L^d} \sum_\vq 
\frac{(\tilde y^{(1)}_{\vk\,\vk-\vq})^2}{(\ep{\vk}+\ep{\vk-\vq})^2}
\frac{ C_d(\vq\sigma)}{[2+(q\xi)^2]^2} ,  \label{avnuk}\\
\av{\pi}_{\vk}  & = V^2 \frac{\sigma^d}{L^d} \sum_\vq
\left \{   
\frac{\tilde y^{(2)}_{\vk\,\vk-\vq}}{2 g \nc\ep{\vk} } - 
\frac{\tilde y^{(1)}_{\vk\,\vk-\vq}\tilde w^{(1)}_{\vk\,\vk-\vq} }{\ep{\vk}(\ep{\vk}+\ep{\vk-\vq})}
\right\} 
\frac{ C_d(\vq\sigma)}{[2+(q\xi)^2]^2} ,  \label{avpik}
\end{eqnarray} 
in terms of the envelopes
\eqref{tildew1}, \eqref{tildey1}, and \eqref{tildey2}. 
In the TF regime, they tend toward  the universal limiting expressions
(dashed black in the insets of Fig.~\ref{nk2.fig}) 
\begin{eqnarray} 
\av{\nu}_{\vk \mathrm{TF}} =\frac{v^2}{4}\frac{[1+(k\xi)^2]^2}{[2+(k\xi)^2]^2}, \label{avnu2k0TF.eq}
\\
\av{\pi}_{\vk \mathrm{TF}} = \frac{v^2}{4}
 \frac{2+4 (k\xi)^2+(k\xi)^4}{[2+(k\xi)^2]^2}, \label{avpi2k0TF.eq}
\end{eqnarray} 
and this independently of the potential details.

\section{Quantum depletion of the condensate} 
\label{condep.sec}

The total particle density $\rho(0) = n  = \nc+\delta n$ is the sum of
condensate density $\nc$ and non-condensed density $\delta
n$. 
The condensate fraction is $\nc/n =1 -\delta n/n$. Within Bogoliubov
theory, the existence of a finite non-condensed fraction $\delta n/n$ at
temperature $T=0$ is called ``quantum
depletion'', because it arises from quantum fluctuations around the
mean-field approximation to the true condensate. From a many-body
point of view, the non-condensed fraction is of course not more quantum than the
condensed one, or perhaps even rather less. 
Here, we follow the established nomenclature and continue to speak of quantum
depletion, at zero temperature, as opposed to the thermal depletion at
finite temperature.

From definitions  \eqref{rhos} and \eqref{nkdef} it follows that
the depleted density $\delta n = n - \nc  $ is the integral of the
fluctuation momentum
distribution,  
\be\label{delnsumk}
\delta n = L^{-d} \sum_\vk \delta n_\vk. 
\ee

\subsection{Homogeneous system} 
\label{homdep.sec}

Let us first recall the homogeneous case $V=0$ \cite{Pitaevskii2003}. Since condensation occurs in the $\vk=0$
mode, the depleted density simply contains all particles with finite
momenta, and the zero-temperature momentum distribution \eqref{delnk0} 
implies 
\be\label{deln0}
\delta n^{(0)} = L^{-d}\sum_{\vk\neq 0} v_k^2 . 
\ee 

In $d=3$, Eq.~\eqref{deln0} evaluates to the depleted density 
$\delta n^{(0)} = [6\sqrt{2}\pi^2 \xi^3]^{-1}$ in the thermodynamic
limit. Equivalently, 
the relative depletion reads 
$\delta n^{(0)}/n = 8 (n a_s^3)^{1/2} / 3\pi^{1/2}$ 
because $g=4\pi\hbar^2a_s/m$ in terms of the
s-wave scattering length $a_s$ and $\xi^2 =
\hbar^2/(2mgn)$ (to leading order, we can identify
$\nc\approx n$ in all perturbative results).  The Bogoliubov ansatz is justified whenever the
fractional depletion is small, $\delta n^{(0)}\ll n$ or equivalently
$n\xi^d\gg1$. This is the case whenever
the so-called gas parameter $na_s^3$ is small, i.e., for low enough
density or weak scattering. 

In $d=2$, one finds 
$\delta n^{(0)} = [8\pi \xi^2]^{-1}$, which is also the result of diagrammatic theory for hard-core
  bosons 
\cite{Schick1971}.
The quantum depletion $\delta n^{(0)}/n$ is roughly independent of density, and
requires weak scattering. 

In $d=1$, the infrared 
$k^{-1}$-divergence of $v_k$ under the integral prevents the existence
of a homogeneous 1D condensate. 
In small enough systems, however, and at very low
temperature, phase fluctuations remain small, and quasi-condensates have
all the attributes of a true condensate 
 \cite{Popov1972,
  Andersen2002,  
  Mora2003,      
  Petrov2000_1D, 
  Fontanesi2010
 }. 
Presently, we are interested in the effect of an external potential on
the homogeneous situation. So we resort to cutting off the 1D-integral at
some value $\aIR=\xi k_\mathrm{IR}\ll1$, with $k_\mathrm{IR}$ of the order
of the inverse system size, and  find
$\delta n^{(0)} = (2\ln2-2-\ln\aIR)/(2\sqrt{2}\pi \xi) $, up to order
$\aIR$. Bogoliubov theory then is valid whenever $n\xi\gg1$, i.e.,
requires high enough 
density in order for the mean-field picture to apply in the first
place. 

So in all relevant dimensionalities, there is a window of validity for
Bogoliubov theory, and the depleted density can be
written $\delta n^{(0)} =
c_d\xi^{-d}$, with a $d$-dependent numerical constant $c_d$ of order
unity or smaller.

\subsection{Potential depletion} 

 In an inhomogeneous system, the quantum depletion cannot be
 calculated by counting all particles with finite momentum, as argued in Sec.~\eqref{condef.sec} above. Instead,  the depleted
density \eqref{delnsumk} is the integral of the fluctuation momentum distribution
\eqref{deltank}, which splits into two contributions: the quantum
depletion of the homogeneous system plus the potential-induced
depletion properly speaking. 
The external potential can change the condensate
fraction because it modifies the local particle
density.  This change in particle density changes the
local interaction energy, which in turn changes the depletion. Since
the interacting system has a nonlinear response, even
in a purely sinusoidal lattice potential the high-density regions will deplete more
condensate than the low-density regions can gain back. At the end, the
presence of the potential causes a net additional depletion of the condensate,
an effect that we propose to call ``potential depletion''.

To our knowledge, the potential depletion, beyond the
mean-field deformation of the condensate, has never been calculated
analytically. In approaches very similar to ours, Singh and Rokhsar 
\cite{Singh1994} arrived at numerical results for the
potential depletion; Lee
and Gunn \cite{Lee1990} estimated a different depletion.  
Within our inhomogeneous Bogoliubov theory, computing the potential
depletion is straightforward: 
Using the perturbative result \eqref{deltan2MV} in \eqref{delnsumk},
we find for the potential-depleted density
\be\label{potdep}
\delta n^{(2)} =  L^{-d} \sum_{\vk,\vq}\tilde M^{(2)}_{\vk\,\vk-\vq}
|\tilde V_\vq|^2 =  \frac{1}{\xi^d(g\nc)^2} \sum_{\vq} M_d(q\xi) |V_\vq|^2.
\ee
The sum over $\vk$ can be carried out without touching the
potential, whence the second equality, which defines the isotropic
depletion kernel for the bare potential,  
\be
M_d(q\xi) = \frac{(\xi/L)^d }{[2+(q\xi)^2]^{2}}\sum_\vk \tilde
M^{(2)}_{\vk\,\vk-\vq}.
\ee     
Prefactors are chosen such that in the thermodynamic limit $M_d(q\xi)$ is a dimensionless function
of $q\xi$ only.

\subsubsection{Lattice potential --} For the lattice potential
\eqref{latticek}, the potential depletion  \eqref{potdep} reads 
\be\label{potdeplattice}
\delta n^{(2)} = \frac{1}{2\xi^d(g\nc)^2} \sum_{j=1}^d V_j^2M_d(K_j\xi) .
\ee
For the 2D lattice potential of Figs.~\ref{figLattice} and
\ref{latticefluctplot}, one finds
$\delta n^{(2)}  =  M_2(K\xi) V^2/(\xi g\nc)^2
\approx 0.141\, \delta n^{(0)}$; for these parameters, 
the potential depletion amounts to only 14\% of the homogeneous depletion. 
These results hold for weak lattices. In much deeper lattices, a tight-binding description becomes more appropriate
 \cite{Xu2006,Oosten2001,Orso2006}.

Let us furthermore check \eqref{potdeplattice} against the 
QMC results of Astrakharchik and
Krutitsky \cite{Astrakharchik2011}, who investigated two different interaction
strengths in a square lattice, such that $K\xi =
6.275 $, and  $K\xi=  1.984$, respectively. 
For these values, \eqref{potdeplattice}  predicts a potential depletion of 
$\delta n^{(2)} =1.05 \, \delta n^{(0)}$, and $0.35 \, \delta
n^{(0)}$, respectively. 
If one takes the QMC values 
for the homogeneous depletion as the reference, then the final
condensate fraction should be $N_0/N = 1-\delta n/n = 0.98$ in the one case and
$0.73$ in the other,
which is in good agreement with the data \cite{Astrakharchik2011}, 
$N_0/N \approx 0.99$ and $0.7$, respectively. Incidentally, the Bogoliubov
prediction for the clean depletion $\delta n^{(0)}/n$  does not
agree so well with the data, which may be due to finite size effects or the slightly
different interaction potential (hard-core
bosons instead of s-wave scattering) used in the QMC approach.

\subsubsection{Random potential --}
\begin{figure}
\hfill\includegraphics[angle=270,width=0.75\linewidth]{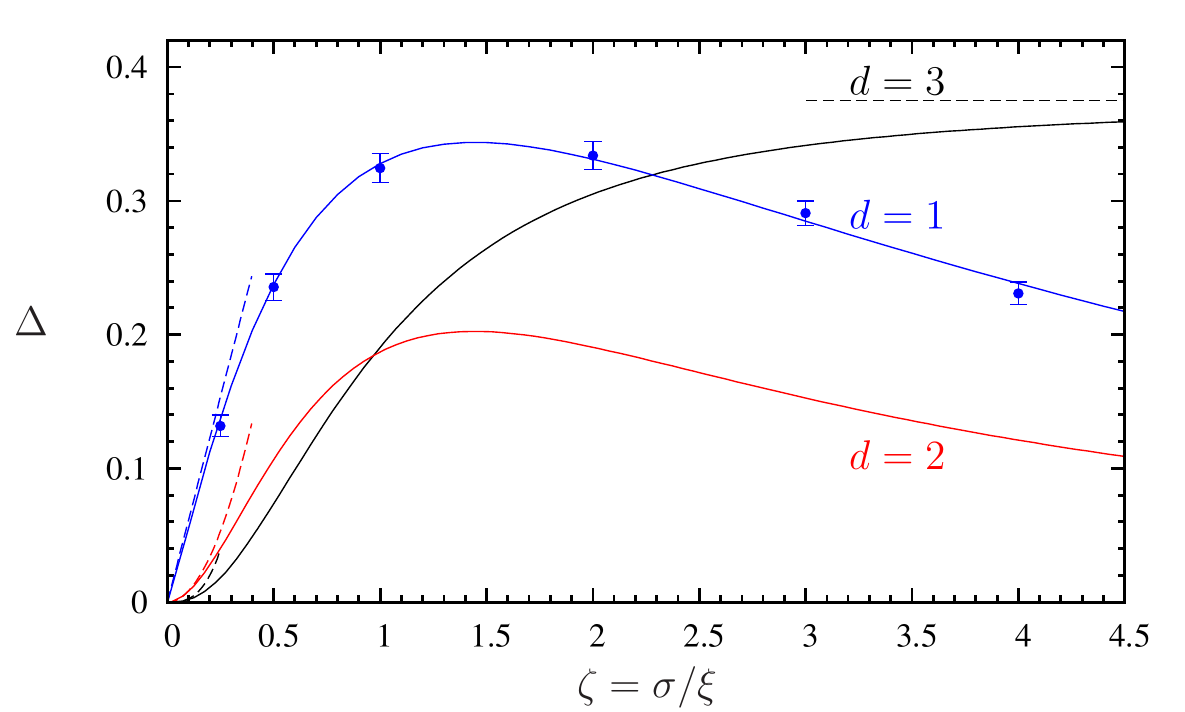}
\caption{
Disorder-induced quantum depletion~\eqref{Deltazetadef}, relative to the clean value and
in units of disorder strength $ v^2$, as function of
the correlation ratio $\zeta=\sigma/\xi$ for the Gaussian
correlation \eqref{gausscorr.eq}.
The curve for $d=1$ depends weakly on the
infrared cutoff $\alpha=\xi k_\mathrm{IR}$ that regularizes already the clean
case; 
$\alpha=0.025$ 
in this plot. Data points are obtained by an exact
numerical diagonalization of the Bogoliubov Hamiltonian
\eqref{eqInhomBgHamiltonian_Gamma}
for $V=0.05 g\nc$, followed by an
ensemble average over disorder. Error bars denote the estimated error
for the average; the number $M$ of realizations is chosen for each
point such that $ML/\sigma =4000$.  
Dashed: 
 universal Thomas-Fermi limit  $\Delta_\mathrm{TF}=3/8$ for very smooth disorder
($\zeta\to\infty$) in $d=3$. The curves for lower dimensions tend to  $\Delta_\mathrm{TF}=0$
and $\Delta_\mathrm{TF}=-1/8$, in $d=2$ and $d=1$, respectively. Dotted: 
limiting behaviour for $\delta$-correlated disorder ($\zeta\ll1$):
$\Delta(\zeta)= \beta_d \zeta^d C_d(0) $ with $\beta_1\approx 0.245$ (for $\alpha=0.05$), $\beta_2\approx 0.135$,
$\beta_3\approx 0.160$. } 
\label{deln_eta_0gaussd123.fig} 
\end{figure} 

Using the momentum distribution \eqref{avdelnk} in \eqref{delnsumk},
the quantum depletion \eqref{potdep}  by a random potential with correlation
\eqref{eqDisorderCorrelator} is found to be  
\be\label{potdepdis}
 \avdn{2}=  \frac{V^2\sigma^d}{(g\nc)^2\xi^dL^d}  \sum_{\vq} 
M_d(q\xi) C_d(\vq\sigma) . 
\ee
Scaled by the homogeneous depletion in the thermodynamic limit, this
can be written as 
\be\label{Deltazeta}
 \avdn{2}=   
v^2\delta n^{(0)} \Delta(\zeta) 
\ee
where $v= V/(g\nc)$ and  
\be\label{Deltazetadef}
\Delta(\zeta)  = \frac{\zeta^d}{c_d} \int \frac{\rmd^d u}{(2\pi)^d} 
M_d(u) C_d(\vu\zeta). 
\ee
This relative potential depletion is found to
be a function of the correlation ratio $\zeta=\sigma/\xi$. Only in $d=1$, it depends also very weakly on the cutoff $\aIR$ that 
 regularizes already the clean depletion.
The convergent integrals \eqref{Deltazeta} require
  \emph{no additional ad-hoc cutoffs},
neither infrared (since the excitations are orthogonal to the
vacuum) nor ultraviolet (since potential correlations
are included). 
Fig.~\ref{deln_eta_0gaussd123.fig} shows $\Delta(\zeta)$ for a
Gaussian-correlated random potential \eqref{gausscorr.eq} in 
dimensions $d=1,2,3$. Data points result from the numerical diagonalization of the Bogoliubov Hamiltonian
\eqref{eqInhomBgHamiltonian_Gamma} in a system of linear size $L$ such
that $\alpha=\xi/L=0.05$, followed by an
ensemble average over disorder. The exact shape of the curve depends on the
correlation function, the general features, however, are rather
robust. 

The asymptotic behaviour of $\Delta(\zeta)$ for very small or very
large correlation lengths is simple.  
In the 
$\delta$-correlated limit $\zeta\to0$ of a white-noise potential, the generic
scaling of \eqref{Deltazetadef} is $\Delta(\zeta) = \beta_d \zeta^d C_d(0)$ with 
$\beta_d=c_d^{-1}  \int \frac{\rmd^d u}{(2\pi)^d} M_d(u)$. 
The numerical coefficients are $\beta_1\approx 0.245$ (weakly dependent on the
cutoff $\alpha$), $\beta_2\approx 0.135$,
$\beta_3\approx 0.160$. In this white-noise regime, the depletion  
depends on $\sigma^dC_d(0)$, and thus requires the existence of a microscopic
correlation scale. In the opposite limit $\zeta\to\infty$ of the Thomas-Fermi
regime, 
the result converges to a truly universal limit
\be\label{DeltaTF}
\Delta_\mathrm{TF} =  c_d^{-1} M_d(0) =
\frac{1}{v^2\delta n^{(0)}}\intdd{k}\avdeln{2}{\vk\mathrm{TF}}
\ee
and evaluates, using \eqref{delnkTF}, to  $\Delta_\mathrm{TF}=3/8$ in
$d=3$ and 
$\Delta_\mathrm{TF}=0$ in $d=2$. In 2 and 3 dimensions,  the
depletion is non-negative, as one would
expect for a random potential that should broaden the momentum
distribution overall. 
For $d=1$ the TF-limit evaluates to the negative value $\Delta_\mathrm{TF}=-1/8$ (in the limit of
infinite system size). This would seem to imply that the random
potential re-populates the condensate.  
But also in $d=1$ the
depletion is positive for most values of $\zeta$, as shown in Fig.~\ref{deln_eta_0gaussd123.fig}. The curve only
crosses over to negative values for such a large value $\zeta=\sigma/\xi$
(depending on the cutoff $\alpha$), that the correlation length $\sigma$
has to be comparable to the system size, which is not the regime of
present interest.  

Interestingly, the TF limit for the potential depletion can also be
derived by the local-density
approximation (LDA) $n_\mathrm{TF} =\nc-V(\vr)/g$ combined with the scaling $\delta n = c_d\xi^{-d} = c'_d \nc^{d/2}$
of the homogeneous depletion 
(Sec.~\ref{homdep.sec}): 

\[ 
\av{\delta n_\mathrm{TF}}  = c_d'n_c^{d/2}\av{[1-V(\vr)/g\nc]^{d/2}} = \delta n^{(0)} \left[1
+ \frac{d(d-2)}{8}v^2 +O(v^3)\right]  
\] 
and thus $\Delta_\mathrm{TF} = d(d-2)/8$, in agreement with the result
of \eqref{DeltaTF}. This argument shows that in $d=2$ dimensions, the TF potential depletion is
zero even non-perturbatively since $\av{V}=0$ without loss of generality. 
This  LDA reasoning works for the correction of the
depletion, but not for the excitation dispersion relation, where genuine
scattering effects determine corrections to the speed of sound
and density of states \cite{Gaul2011_bogoliubov_long,Gaul2009a}, 
and furthermore cause exponential localization
\cite{Lugan2011,Bilas2006,Lugan2007}.  

Summarizing the results of this section, we conclude that the combined depletion due to
interaction and external potential reads 
\be
\delta n=\delta n^{(0)}[1 + v^2 \Delta],
\ee with
$|\Delta|\ll1$.  
Clearly, the potential depletion alone,   
$\delta n^{(2)}/n = (\delta n^{(0)}/n) v^2 \Delta$, is at least a factor 
$\delta n^{(0)}/ n\ll1$ smaller than the mean-field condensate
deformation \eqref{nck}, which is of order $v^2$.
In hindsight, this result is
rather plausible: the primary effect of the external potential is
merely to deform the condensate. The potential depletion is a
secondary effect, caused by enhanced interaction in the regions of higher density. 
We conclude that, as long as the
original assumption of a non-zero condensate amplitude holds,
our inhomogeneous Bogoliubov theory applies to Bose condensates in rather inhomogeneous potentials.

\section{Summary}
\label{conclusion.sec}

We have investigated the effect of external potentials on 
Bose-condensed gases using inhomogeneous Bogoliubov theory. The
principal effect of an
external potential is to deform the mean-field condensate. 
Secondly, the potential affects the momentum distribution of quantum
fluctuations, 
for which we have obtained a general expression.   
Finally, we have calculated the quantum depletion induced by the
external potential, or potential depletion for short. 
In detail, we have
studied lattices and spatially correlated random potentials. The potential
depletion turns out to be proportional to the
homogeneous depletion, a fact that underscores the 
applicability of inhomogeneous Bogoliubov theory in weak to moderately strong
potentials.  Our
analytical predictions are in agreement with a numerical diagonalization
of the Bogoliubov Hamiltonian as well as with recent quantum Monte Carlo
simulations \cite{Astrakharchik2011}. The  
inhomogeneous Bogoliubov theory shown at work here is therefore proven
capable of describing the excitations of weakly interacting
condensates in external potentials, and from there ought to provide
many of other static
and dynamic properties.    

\ack 

This work is supported by the National Research Foundation \& Ministry of
Education, Singapore. 
Research by C.G. was supported by a PICATA postdoctoral fellowship of the Moncloa Campus of International Excellence (UCM-UPM). We have benefitted from discussions with
G.~Astrakharchik,  L.~Fontanesi, P.~Lugan, A.~Pelster, and I.~Zapata.  

\appendix 
\section{Transformation to Bogoliubov quasi-particles}
\label{Bogotrafo.apx}

The  matrices  $\ukp$ and $\vkp$ as defined by \eqref{ukp} and
\eqref{vkp} 
are the Fourier components with wave
vector $\vk$ of the modes $u_\vp(\vr)$
and $v_\vp(\vr)$ defined in \cite{Gaul2011_bogoliubov_long}. As
explained there, the momentum index $\vp$ can be used to label the modes even in the inhomogeneous
setting. The transformation
\eqref{delPsikgamma} preserves the canonical commutation 
relation, and thus guarantees 
$[\gh{\vp},\ghd{\vp'}]=\delta_{\vp\vp'}$ as well as $[\gh{\vp},\gh{\vp'}]=0$, via the completeness relations
\begin{eqnarray}
\sum_\vp \left(u_{\vk\vp}u_{\vk'\vp}^* -\vkp v_{\vk'\vp}^*\right) =
  \delta_{\vk\vk'}, \label{ortho1a}\\
 \sum_\vp \left(u_{\vk\vp}v_{\vk'\vp}^* -\vkp u_{\vk'\vp}^*\right) =
 0.\label{ortho1b}
\end{eqnarray}
The non-symmetric matrices $\ukp\neq u_{\vp\vk}$ and $\vkp\neq
v_{\vp\vk}$ 
also satisfy the biorthogonality
\begin{eqnarray}
\sum_\vk(\ukp\ukpr^* - \vkp\vkpr^*) = \delta_{\vp\vp'}, \label{ortho2a}\\
\sum_\vk(\ukp\vkpr^* - \vkp\ukpr^*) = 0.  \label{ortho2b}
\end{eqnarray} 

The zero mode deserves special attention because 
$a_p = (\epn{p}/\ep{p})^{1/2}$ diverges as $p^{-1}$ when $p\to
0$. In this range, elementary excitations are essentially phase fluctuations. 
Setting $a_0=1$, one finds that
the Bogoliubov excitation
\be\label{gamma0def}
\gh{0} = \delta\hat n_0/(2\nc^{1/2})  + \rmi \nc^{1/2} \delta
\hat\varphi_0, 
\ee
together with its Hermitian conjugate, describes the number fluctuation 
\begin{equation}\label{Goldstone}
 \delta \hat n_0 =  \sqrt{\nc} (\gh{0} + \ghd{0}) 
= L^{-\frac{d}{2}} \int \rmd^dr \Phi(\vr)\left[\delta
  \hat\Psi(\vr)+\delta \hat\Psi^\dagger(\vr) \right]. 
\end{equation}
This operator, called $\hat P$ in
\cite{Lewenstein1996}, generates an exact zero-energy (Goldstone) mode of the
U(1)-symmetry breaking Bose condensed state. 
The corresponding mode functions 
are
\be
u_{\vk0} =[\Phi_\vk + \check\Phi_\vk]/(2N_c^{1/2}), \qquad
v_{\vk0} =[\Phi_\vk - \check\Phi_\vk]/(2N_c^{1/2}). 
\ee 
With these definitions, the completeness relations \eqref{ortho1a}-\eqref{ortho1b} and
biorthogonality relations \eqref{ortho2a}-\eqref{ortho2b} include and extend to  
the zero modes. 
In the present article, we investigate the spatial
structure of quantum fluctuations, and
 the contribution from $\vp=0$ has vanishing
weight anyway in the thermodynamic limit where sums over momenta
turn into integrals---except for 1D, but there, we introduce an IR cutoff to
regularize the divergence. This masks the phase diffusion physics
at long distances and times
\cite{Lewenstein1996}, which has not been
the subject of the present investigation, but would certainly be
  worthwhile studying in greater detail 
\cite{Mora2003,Petrov2000_1D,Fontanesi2010}.  

\bibliographystyle{mybst_noeprint_notitle}
\section*{References}
\bibliography{literatureBEC}

\end{document}